\documentclass[%
reprint,
superscriptaddress,
 amsmath,amssymb,
 aps,
pra,
]{revtex4-2}

\usepackage{graphicx}
\usepackage{dcolumn}
\usepackage{bm}
\usepackage{amsmath}
\usepackage{systeme}
\usepackage{dsfont}
\usepackage{tcolorbox}
\usepackage{tabularx}
\usepackage{colortbl}
\tcbuselibrary{skins}

\usepackage[colorlinks, linkcolor=red, anchorcolor=blue, urlcolor=blue, citecolor=blue]{hyperref}

\newcolumntype{Y}{>{\centering\arraybackslash}X}
\tcbset{tab2/.style={enhanced, fonttitle=\bfseries,fontupper=\smallsize\sffamily,
colback=grey!2,colframe=black,colbacktitle=yellow!40!white,
coltitle=black,center title}}

\newcommand{\beq}{\begin{equation}}
\newcommand{\eeq}{\end{equation}}
\newcommand{\beqa}{\begin{eqnarray}}
\newcommand{\eeqa}{\end{eqnarray}}

\begin{document}

\title{Fast transport and splitting of spin-orbit-coupled spin-1 Bose-Einstein Condensates}

\author{Yaning Xu}
\affiliation{Institute for Quantum Science and Technology, Department of Physics, Shanghai University, Shanghai 200444, China}

\author{Yuanyuan Chen}
\email{cyyuan@shu.edu.cn}
\affiliation{Institute for Quantum Science and Technology, Department of Physics, Shanghai University, Shanghai 200444, China}

\author{Xi Chen}
\email{xi.chen@ehu.eus}
\affiliation{Department of Physical Chemistry, University of the Basque Country UPV/EHU, Apartado 644, 48080 Bilbao, Spain}	
\affiliation{EHU Quantum Center, University of the Basque Country UPV/EHU, 48940 Leioa, Spain}

\begin{abstract}

In this study, we investigate the dynamics of tunable spin-orbit-coupled spin-1 Bose-Einstein condensates  confined within a harmonic trap, focusing on rapid transport, spin manipulation, and splitting dynamics. Using shortcuts to adiabaticity, we design time-dependent trap trajectories and spin-orbit-coupling strength to facilitate fast transport with simultaneous spin flip. Additionally, we showcase the creation of spin-dependent coherent states via engineering the spin-orbit-coupling strength. To deepen our understanding, we elucidate non-adiabatic transport and associated spin dynamics, contrasting them with simple scenarios characterized by constant spin-orbit coupling and trap velocity. Furthermore, we discuss  the transverse Zeeman potential and nonlinear effect induced by interatomic interactions using the Gross-Pitaevskii equation, highlighting the stability and feasibility of the proposed protocols for the state-of-the-art experiments with cold atoms.
  
\end{abstract}

\maketitle 

\section{INTRODUCTION}

Spin-orbit coupling (SOC) denotes the interplay between the spin and momentum of a quantum particle, holding relevance for numerous fundamental applications in various quantum systems, such as semiconductor quantum dots \cite{RevModPhysspintronics},  topological superconductors \cite{Phystoday}, and ultracold atomic gases \cite{lin2011spin}. In condensed-matter physics, SOC effects are associated with a large amount of novel quantum phenomena, such as spin Hall effect and topological insulators, see the review \cite{bihlmayer2022rashba}. Notably, by exploiting the coupling between electron spin and its motion, spin qubits with long coherence time are essential for quantum computation and information processing tasks \cite{Losspra,ReviewDaniel}. Understanding and controlling SOC within nanometer-scale semiconductor devices are crucial steps toward realizing efficient and reliable quantum technologies \cite{RevModPhysspinqubit}. In particular, semiconductor nanowire quantum dots with strong SOC can be harnessed to manipulate a spin-orbit qubit, employing techniques like electric-dipole spin resonance \cite{YouPRL} and time-dependent modulation \cite{YuePRL,Ramsakprl,PRBsherman}. Along with this, the controllable confinement potential or/and SOC can be further exploited to generate non-classical states \cite{Bednarekprb94,Bednarekprb96} and quantum gates \cite{Li_2018} within semiconductor quantum dots.

In a similar vein, the experimental realization of synthetic SOC of ultra-cold atomic systems \cite{lin2011spin,galitski2013spin} has recently emerged as a powerful tool for manipulating the internal freedom of atoms and exploring quantum states \cite{Zhai_2015}.  The characteristics associated with tunable SOC effects, such as collective modes, novel dynamics of the orbit and spin, negative-mass hydrodynamics, have been subject to extensive investigation \cite{Panprlcollective,SpielmanPRLtunableSOC,Benjaminprl,Stringaripra12,Chuanweipra,Anpra16,Negativemassprl}, yielding fruitful SOC-related physics with ultracold atoms and potential applications in ``atomtronics" \cite{ZBpra13,PhysRevALZ,Shermanpra,detlapra20,mossman2019experimental,li2019spin}. While electrons primarily manifest as spin-1/2 systems, neutral atoms, boasting an abundance of hyperfine states, can exhibit higher pseudospins. 
Through techniques such as coupling three hyperfine states with Raman lasers, for instance, spin-orbit coupled spin-1 Bose-Einstein condensates (BECs) can be experimentally constructed in a gas of ${ }^{87} \mathrm{Rb}$ \cite{Campbell2016magnetic,luo2016tunable}. Such systems, with a richer spin structure compared to spin-orbit coupled spin-1/2 BECs, can exhibit multiple internal spin states, allowing for more complex spin-mixing dynamics, collective excitations, and topological excitations \cite{Patrikpra,Colepra15,Chuanweipra16,Stringariprl16,PhysRevA.106.013304}. Particularly,  spin-orbit-coupled spin-1 BECs can undergo quantum phase transitions between different ground state phases, including plane-wave, stripe, and zero-momentum phases, driven by changes in external parameters such as magnetic fields or interaction strengths \cite{PhysRevA.103.L011301,PhysRevA.104.L031305,chen2022elementary,PhysRevA.108.043310}. Accordingly, such exotic systems enable enhanced control of the center of mass \cite{Yongpingpra21,Wenxianpra}, and also facilitates the manipulation of spin dynamics driven by shaking harmonic traps, as well as engineering SOC strength and Zeeman fields \cite{Xiinversepra,Suotangpra19,SuotangPRA2020}.

In this paper, our goal is to manipulate spin dynamics and orbital motion to achieve fast transport and splitting of tunable spin-orbit-coupled spin-1 BECs confined within a harmonic trap. Based on single-particle Hamiltonian, precise control over the spin state and orbital motion is made possible by the analytical solvability of the exact wave function,  especially in the absence of atomic interaction and transverse potential. \cite{cadez_2013,Nonadiabaticprl14}. To realize this objective, we utilize inverse engineering \cite{Xiinversepra}, a method inspired by shortcuts to adiabaticity (STA) \cite{RevModPhys.91.045001}, to design time-dependent SOC strength and/or trap trajectories. This approach enables fast transport with simultaneous spin flips and the splitting of wave packets within short time scales as well. Additionally, we compare the outcomes of simple adiabatic scenarios with constant SOC strength and trap velocity. Furthermore, we delve into the stability analysis against the transverse-Zeeman potential and nonlinear effect, through comprehensive numerical simulations based on the Gross-Pitaevskii equation. Although our primary focus is on BECs, the results also have implications for other systems such as quantum dots \cite{Bednarekprb94,Bednarekprb96,Li_2018}.

The remainder of this paper is organized as follows: In Sec. \ref{Preleminaries}, we introduce the Hamiltonian and model, and present the analytical solution for cold atoms trapped in a harmonic trap by neglecting atomic interactions. Building upon this, Sec. \ref{engineering} details the inverse engineering method for achieving fast transport with spin flip and splitting using the designed SOC strength and/or trap trajectory. Subsequently, in Sec. \ref{constant}, we compare these results with the case of constant SOC strength and trap velocity. Additionally, we check the stability against the nonlinearity induced by interatomic interactions for different initial states and transverse Zeeman potential in Sec. \ref{nonlinear}. Finally, a brief conclusion is provided in Sec. \ref{conclusion}.

\section{Preliminaries}

\label{Preleminaries}

We begin with spin-1 BECs confined in a one-dimensional harmonic trap along the $x$-axis with SOC. The experimental setup for Raman-induced SOC in a spin-1 BEC is detailed in Ref. \cite{Campbell2016magnetic},
where a bias magnetic field along a specific axis, typically the $z$-axis, to manipulate the atomic states and induce Zeeman splitting, and a pair of counter-propagating Raman lasers with orthogonally polarized beams are further utilized to coherently couple the different hyperfine states of the $|F=1, m=0, \pm 1 \rangle$ manifold, thereby inducing SOC.
In the context of a one-dimensional harmonic potential, the system can be simply described by the Gross-Pitaevskii equation (GPE), 
\begin{equation}
\label{GP}
i \hbar \frac{\partial \Psi(x, t)}{\partial t} = (H  + H_{int}) \Psi(x, t),
\end{equation}
where the spinor wave function $\Psi(x, t) =\left[\psi_{1}, \psi_{0},\psi_{-1}\right]^{\mbox{T}}$ describes the occupation of three components. By neglecting a dispensable constant, 
the single-particle Hamiltonian in Eq. (\ref{GP}) is written as  \cite{Patrikpra,Colepra15,Chuanweipra16,Stringariprl16,SuotangPRA2020}
\begin{equation}
\label{h}
H = \frac{p^2_x}{2m} + \frac{1}{2}m\omega^2 [x-x_0(t)]^2 + \alpha(t) p_x F_z +  \hbar \Omega F_x.
\end{equation}
Here, $m$ is the mass of the atom, $p_x$ denotes the momentum operator along the $x$ direction, $F_z$ corresponds to the $z$-component of the $3\times 3$ spin-1 Pauli matrices. The external trap
is moving harmonic potential, with time-dependent trajectory $x_0(t)$ of trap center and fixed trap frequency $\omega$. The term $\propto p_x F_z$  stands for the Raman-induced SOC with tunable strength $\alpha(t)$, achievable through the modulation of external magnetic fields \cite{Campbell2016magnetic,luo2016tunable}.  Also, $\Omega= \Omega_R/\sqrt{2}$ is the transverse-Zeeman potential, with $\Omega_R$ being the Raman Rabi frequency associated with the Raman laser process. However, 
the detuning $\delta$ can indeed be utilized to select two out of the three Zeeman states as a spin 1/2 system \cite{lin2011spin,Panprlcollective,PhysRevALZ}. In this context, we set $\delta=0$ within the $F_z$ term to maintain Raman resonance.
In the GPE, the nonlinear term $H_{int}$, depending on the number of atoms $N$, is described by the density-density interaction with the coefficient  $c_0$ and the spin-spin interaction with the coefficient $c_2$, as we will discuss later. As a reminder, the entire wave function satisfies the normalization condition, 
$\int dx (|\psi_{1}|^2+|\psi_{0}|^2+|\psi_{-1}|^2)=N$ (equals to $=1$ in the absence of nonlinear interaction).

In the actual setup, the spin-orbited coupled BEC was realized in a harmonic potential with frequency $\omega=2 \pi \times 250 \mathrm{~Hz}$, where ${}^{87} \mathrm {Rb} $ atomic mass is  $m = 1.443 \times10^ {- 25} \mathrm {~ kg}$ \cite{Campbell2016magnetic,luo2016tunable}. In order to simplify the numerical calculation using the split operator method for spinor wave-packet dynamics \cite{Chaves2015,gawryluk2018unified}, we adopt $m=\hbar=\omega=1$, so the units of the relevant physical parameter are relabelled as $T=1/\omega\approx0.637 \mathrm{~ms}$ and the characteristic length $a_0 = \sqrt {\hbar/(m \omega)} \approx0.682 \mathrm {~\mu m} $ correspondingly.  

Regarding the single-particle Hamiltonian \eqref{h}  in absence of the transverse potential $(\Omega=0)$, the exact solution of the time-dependent Schr\"{o}dinger equation can be obtained through a unitary transformation $\mathcal{U}(t)$, such that
\begin{equation}
H_0 =\mathcal{U}(t) H(t) \mathcal{U}^{\dag}(t) - i \dot{\mathcal{U}}(t) {\mathcal{U}}^{\dag} (t)=
\frac{p^2_x}{2m} + \frac{1}{2}m \omega^2 x^2,
\end{equation}
becomes a stationary harmonic oscillator without SOC. Similar to the spin-1/2 case \cite{cadez_2013,Nonadiabaticprl14}, the unitary transformation $\mathcal{U}(t) = \mathcal{U}o(t) \mathcal{U}s(t)$ comprises two parts:
\begin{eqnarray}
\label{U0}
\mathcal{U}_o(t)&=& e^{-\frac{i}{\hbar} \phi_{x_0}(t)} e^{-i x_c(t) p} e^{\frac{i}{\hbar} m \dot{x}_c(t) x},
\\
\label{Us}
\mathcal{U}_s(t) &=& e^{-i \phi_{\alpha}(t)} e^{-i \phi(t) F_z} e^{-\frac{i}{\hbar} m \alpha_c(t) x F_z} e^{-\frac{i}{\hbar \omega^2} \dot{\alpha}_c(t) p F_z},~~~
\end{eqnarray}
where the dot represents the time derivative,  $\phi_{\alpha, {x_0}} (t)$  are two action phase factors defined as $ \phi_{\alpha, {x_0}} (t)=-\frac{1}{\hbar} \int_0^t d \tau  \mathcal{L}_{\alpha,{x_0}}(\tau)$,
with the Lagrange functions $\mathcal{L}_\alpha(t)  = m \dot{\alpha}_c^2(t)/(2 \omega^2)- m \alpha_c^2(t)/2 +m \alpha_c(t) \alpha(t)$,
and $\mathcal{L}_{x_0}(t)  = m \dot{x}_c^2(t)/2- m \omega^2\left[x_c(t)-x_0(t)\right]^2/2$, and the phase factor $\phi (t)$ given by,
\begin{equation}
\label{phi}
\phi (t)=-\frac{m}{\hbar} \int_0^t \dot{\alpha}_c(\tau) x_0(\tau) d \tau.
\end{equation} 
The parameters $\alpha_c(t)$ and $x_c (t)$ satisfy auxiliary equations 
\begin{align}
\label{xc}
&\ddot{x}_c(t)+\omega^2\left[x_c(t)-x_0(t)\right] =0,
\\
\label{alpha}
&\ddot{\alpha}_c(t)+\omega^2\left[\alpha_c(t)-\alpha(t)\right] =0.
\end{align}
In this scenario, the solution of the original Hamiltonian \eqref{h} is expressed as
$|\Psi(t)\rangle=\mathcal{U}^{\dag}(t)|\psi(t)\rangle$, where $|\psi(t)\rangle$ represents the solution of the transformed Hamiltonian $H_0$. Consequently, we have
\begin{equation}
|\Psi(t)\rangle = \mathcal{U}^{\dag}(t) e^{-iH_0t/\hbar} \mathcal{U}(0) |\Psi(0)\rangle.
\end{equation}
In the case where the initial state is an eigenstate of $H(0)$, i.e., $|\Psi_{ns}(0)\rangle = \mathcal{U}^{\dag}(0) |\psi_n\rangle |\chi_s\rangle$, with $|\psi_n\rangle$ denoting the $n$-th eigenstate of the stationary harmonic oscillator $H_0$, and $|\chi_s\rangle$ representing a spinor with spin $s$, the time-evolved state simplifies to:
\begin{equation}
\label{Psi}
|\Psi_{ns}(t)\rangle = e^{-i\omega_nt} \mathcal{U}^{\dag}(t) |\psi_n\rangle |\chi_s\rangle,
\end{equation}
where $\omega_n=(n+1/2)\omega$. In the subsequent discussion, the ground state is of particular interest when $n=0$.

Next, we utilize Eqs. \eqref{phi}-\eqref{alpha} to design fast transport with  spin flip and splitting using inverse engineering. The proposed strategy involves taking the position of the trap $x_0(t)$ and the time-dependent SOC strength $\alpha(t)$ as free parameters first, and then engineering inversely through the solutions of $x_c(t)$ and $\alpha_c(t)$ that meet the specified boundary conditions.

\section{Inverse Engineering} 

\label{engineering}

\subsection{Fast transport with spin flip}

In this section, to manipulate the spin states of atoms and facilitate fast transport, we employ inverse engineering to design the SOC strength $\alpha(t)$ and the trap trajectory $x_0(t)$ according to Eqs. \eqref{xc} and \eqref{alpha} by selecting appropriate boundary conditions. 
Assuming the trap's minimum value starts at $x_0(0) = 0$ and ends at $x_0(t_f) = d$ after a time interval $t_f$, we opt for boundary conditions that ensure accelerated transport without final excitation:
\begin{equation}
\label{BC-1}
\begin{gathered}
x_c(0)=0, \quad  \dot{x}_c(0)=0, \quad  \ddot{x}_c(0)=0,\\
x_c\left(t_f\right)=d,  \quad \dot{x}_c\left(t_f\right)=0,\quad  \ddot{x}_c\left(t_f\right)=0.
\end{gathered}
\end{equation}
Various functions meet these conditions, and for simplicity, we adopt a trigonometric ansatz in the form $x_c(t)=\sum_{n=0}^5 a_n \cos{(n \pi t)} $. Consequently, the center-of-mass of the cold atoms, $x_{c}(t)$, is expressed as:
\begin{equation}
x_c(t)=d\left[\frac{1}{2}-\frac{25}{32}\cos{(3\pi s)}+\frac{9}{32}\cos{(5\pi s)}\right],
\end{equation}
where $s=t/t_f$. Utilizing Eq. \eqref{xc}, the potential trajectory $x_0(t)$ is obtained as $x_0(t)=x_c(t)+\ddot{x}_c/\omega^2$.

\begin{figure}
\centering
\includegraphics[width=0.45\textwidth]{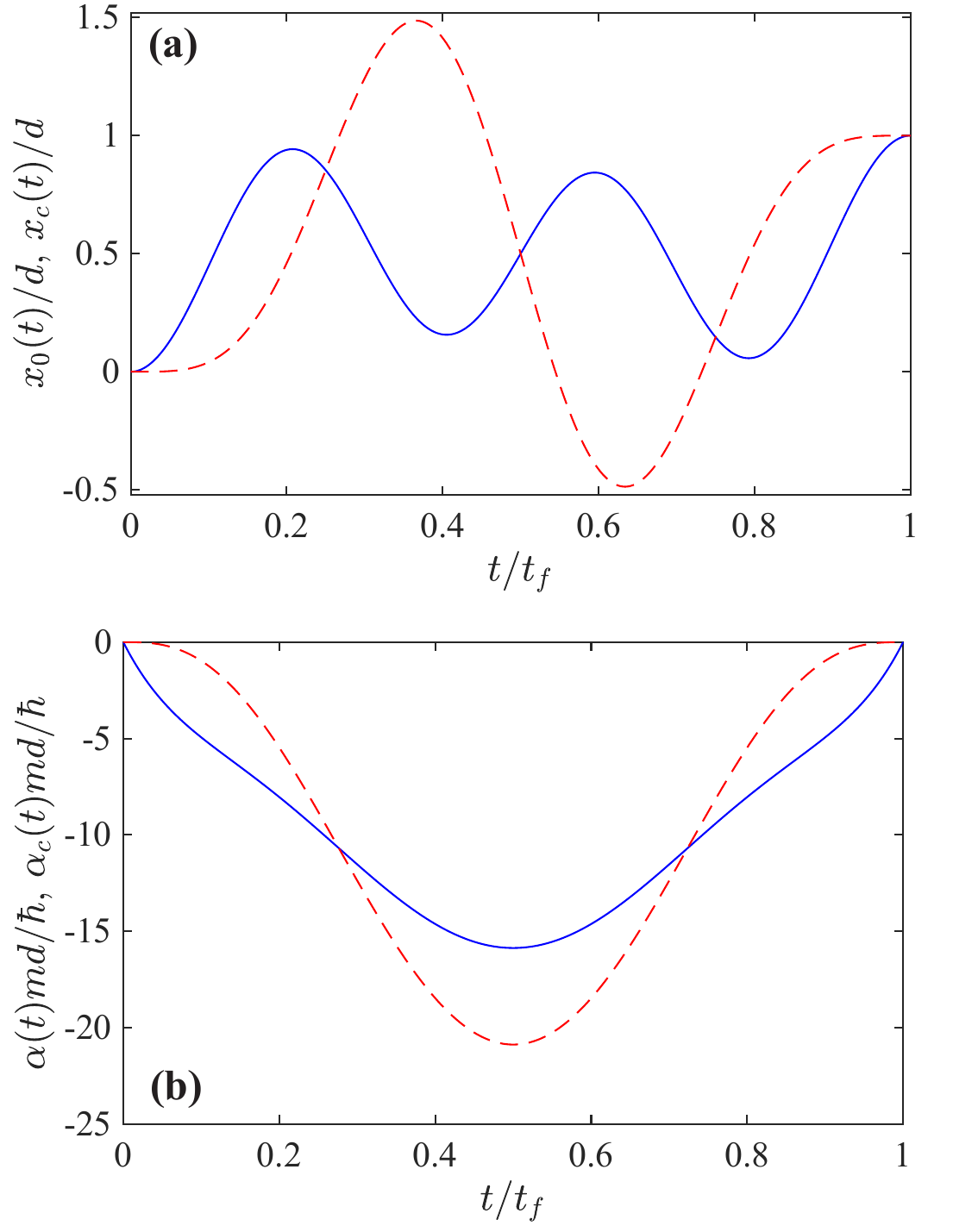}
\caption{(a) Dependence of the trap position $x_0(t)/d$ (solid blue line) and the center of mass $x_c(t)/d$ (dashed red line) of atoms on time $t/t_f$. (b) Dependence of the SOC strength $\alpha(t)md/\hbar$ (solid blue line) and the auxiliary parameter $\alpha_c(t)md/\hbar$ (dashed red line) on time $t/t_f$. All quantities are dimensionless.}
\label{x&alpha}
\end{figure}

In order to control the spin states of atoms, we also design the SOC strength $\alpha(t)$ according to the Eqs. \eqref{phi} and \eqref{alpha} by choosing appropriate boundary conditions. From the expression of unitary, we have to set 
\begin{equation}
\label{BC-2}
\begin{gathered}
\alpha_c(0)= \dot{\alpha}_c(0) = \ddot{\alpha}_c(0)=0,\\
\alpha_c\left(t_f\right)= \dot{\alpha}_c\left(t_f\right)= \ddot{\alpha}_c\left(t_f\right)=0,
\end{gathered}
\end{equation}
and 
\begin{equation}
\phi(t_f)=-\frac{m}{\hbar} \int_0^{t_f} \dot{\alpha}_c(\tau) x_0(\tau) d \tau=\pi,
\end{equation}
yielding the spin flip, that is the spin rotation around the $z$-axis by $\pi$ angle.  Here, we just provide an example of spin flip.  Of course, one can choose another angle for spin rotation or even nonadiabatic holonomic transformations \cite{Ramsakprl}. Instead of trigonometric function, simple and flexible polynomial ansatz can be fitted according to the boundary conditions mentioned above in the form $\alpha_c(t)=\sum_{n=0}^6 a_n t^n $, and  we finally obtain
\begin{equation}
\alpha_c(t) = \frac{\mathcal{N} \hbar}{m d}(s^6-3 s^5+3 s^4-s^3), 
\end{equation}
where
\[
\resizebox{\linewidth}{!}{%
$\displaystyle
\mathcal{N} = \frac{253125\pi^7\omega^2 t_f^2 }{8(418950\pi^2-344250\pi^4 -48008\omega^2t_f^2+41895\pi^2\omega^2t_f^2)}.
$
}
\]
Once $\alpha_c(t)$ is obtained, the SOC strength $\alpha(t)$ can be derived from Eq. \eqref{alpha}, $\alpha(t)=\alpha_c(t)+\ddot{\alpha}_c/\omega^2$. Fig. \ref{x&alpha} (a) illustrates the dimensionless trap trajectory $x_0(t)/d$ (solid blue line) designed from the center-of-mass $x_c (t)/d$ of atoms (dashed red line). Additionally, Fig. \ref{x&alpha} (b) demonstrates the dimensionless SOC strength $\alpha (t) m d/\hbar$ (solid blue line), alongside the evaluation of the parameter $\alpha_c (t)m d/\hbar$ (dashed red line).

Next, we investigate the fast transport of an atomic wave packet using the designed moving trap and SOC strength, as illustrated in Fig. \ref{x&alpha}. Based on Eq. (\ref{Psi}), we assume that the initial state is given by:
\begin{equation}
\label{initial}
|\Psi(x, 0)\rangle=\frac{1}{2}\left(\begin{array}{l}
 1, \
\sqrt{2}, \ 1
\end{array}\right)^{\mbox{T}}  \otimes|\psi(x, 0)\rangle,
\end{equation}
where the ground state is the wave function of the stationary harmonic oscillator, resulting in
\begin{equation}
|\psi(x, 0)\rangle=\left(\frac{1}{\pi a^2}\right)^{1 / 4} \exp \left[-\frac{x^2}{2 a^2}\right].
\end{equation}
The wave function of the final state is given by:
\begin{equation}
\label{final}
|\Psi(x, t_f)\rangle=\frac{1}{2}\left(\begin{array}{l}
 1, \
-\sqrt{2}, \ 1
\end{array}\right)^{\mbox{T}}  \otimes|\psi(x, t_f)\rangle,
\end{equation}
with
\begin{equation}
|\psi(x, t_f)\rangle=\left(\frac{1}{\pi a^2}\right)^{1 / 4} \exp \left[-\frac{(x-d)^2}{2 a^2}\right].
\end{equation}
As an example, our objective is to transport the ground state from $x=0$ to $x=d$ within a shorter time $t_f$ without inducing final excitation, achieved by applying appropriate boundary conditions \eqref{BC-1}. Additionally, the boundary condition $\phi (t_f) = \pi $ in Eq. \eqref{phi} also induces a spin rotation through $\exp{[-i \phi (t) F_z ]}$. Specifically, we assume that the initial spinor function $|\chi_s \rangle$ is one of the eigenstates of $F_x$, namely, $(1/2, 1/\sqrt{2}, 1/2)^{\mbox{T}}$, transferring to the other one, $(1/2, -1/\sqrt{2}, 1/2)^{\mbox{T}}$, at $t=t_f$.

\begin{figure}
\centering
\includegraphics[width=0.45\textwidth]{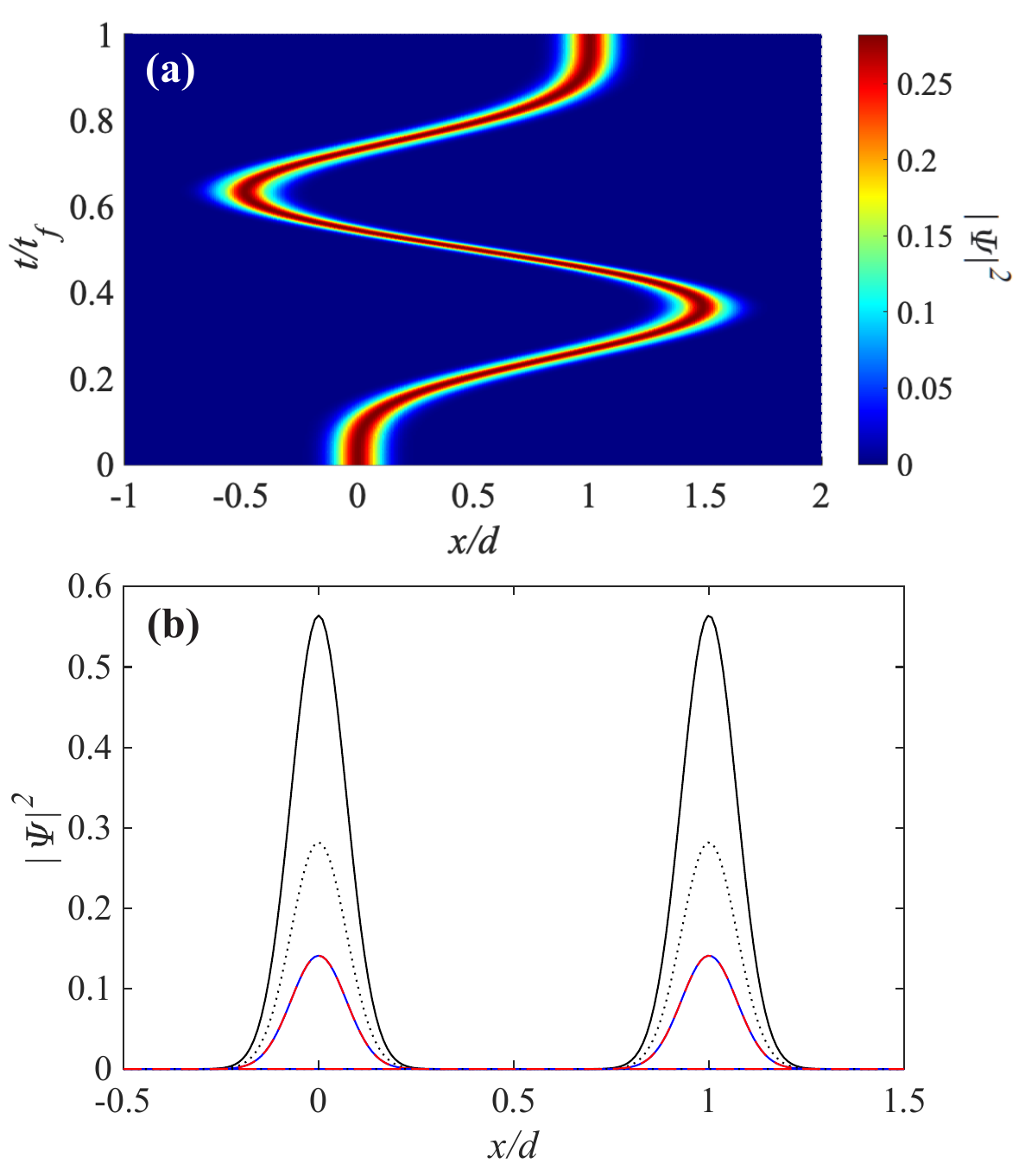}
\caption{(a) Depicts the propagation contour map of wave packets during the fast transport process designed using the inverse engineering method. (b) Illustrates the density distribution of the total wave function $|\Psi(x,t)|^2$ (black line) at $t=0$ and $t=t_f$, along with the density distribution of the three spin components $|\Psi_{1,0,-1}(x,t)|^2$, denoted by blue solid line, black dotted line, and red dashed line, respectively.
The parameters $t_f=10$ and $d=10$ are given in the units of $T$ and $a_0$, respectively, as described in the main text.}
\label{wavepacket}
\end{figure}

Figure \ref{wavepacket} provides insights into the transport of atomic wave packets facilitated by a moving trap and SOC strength designed as shown in Figure \ref{x&alpha}. In panel (a), the propagation of the wave packet during transport is illustrated, highlighting the effectiveness of the designed trap trajectory. Panel (b) compares the initial and final states. Clearly, in accordance with the designed trap trajectory, the spin-1 atomic wave packet is transported from $x_0=0$ to $x_0=d$ without experiencing any excitation. Meanwhile, the probabilities of the three spin components remain equal, due to the condition of spin flip.

 \begin{figure}
\centering
\includegraphics[width=0.45\textwidth]{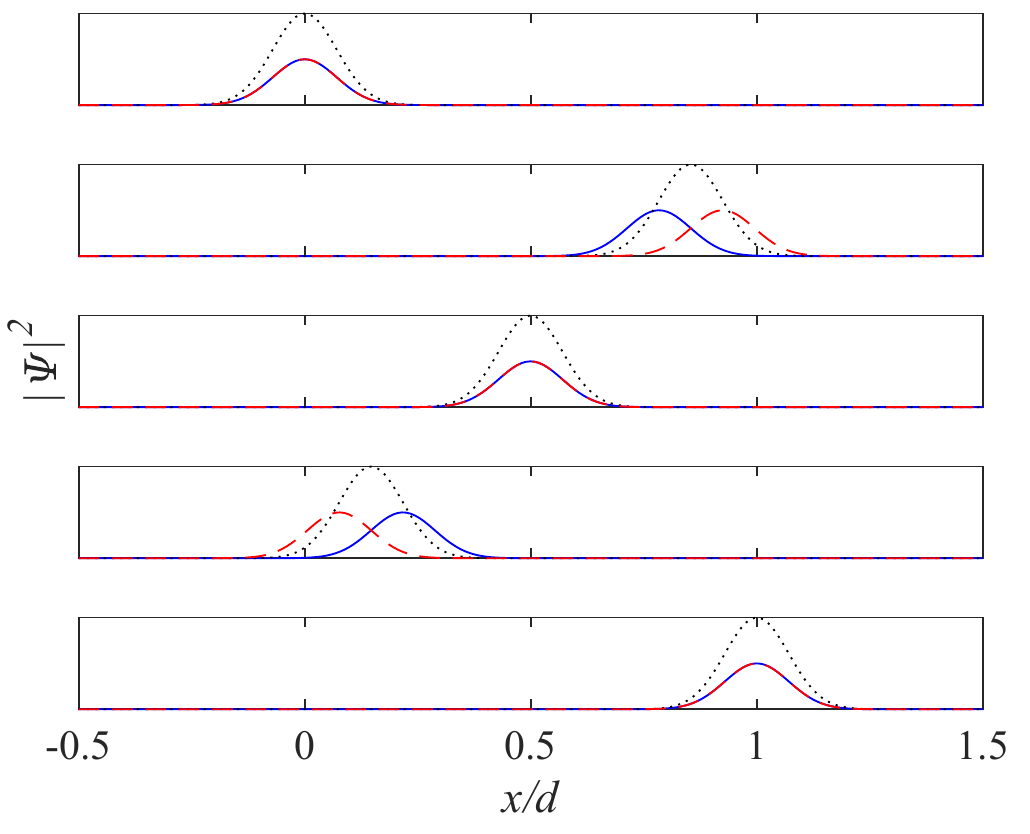}
\caption{ Time evolution of the wave packets with spin-up (solid blue line), spin-zero (dotted black line) and spin-down (dashed red line) components at different times: $ t = 0, t_f /4, t_f /2, 3t_f /4, t_f$. Parameters are the same as those in Fig. \ref{wavepacket}.}
\label{time-evoluation}
\end{figure}



To clarify the detailed spin rotation, we examine the propagation of the three spin components as depicted in Fig. \ref{time-evoluation}. Initially, the spin, not being an eigenstate of the Hamiltonian \eqref{h}, commences rotating, leading to the splitting of the wave packet into three components with distinct velocities. This splitting is elucidated by the spin-dependent contribution, where the velocity operator, 
\begin{equation}
\label{velocity}
v=\frac{i}{\hbar} \left[H,x\right]=\frac{p}{m}+\alpha (t) F_z,
\end{equation}
yields varying velocities for the spin components.  At the beginning, with the spin parallel to the $x$-axis, the expectation value of velocity is $0$, triggering the wave packet components to diverge. At the midpoint, $t = t_f/2$, the spin aligns parallel to the $y$ direction, causing the three spin components to reconverge with an equal but nonzero expectation value of velocity. Ultimately, per boundary condition \eqref{phi}, the spin becomes antiparallel to the $x$-axis, resulting in the wave packet's three spin components coinciding and zero velocity.

\begin{figure}
\centering
\includegraphics[width=0.45\textwidth]{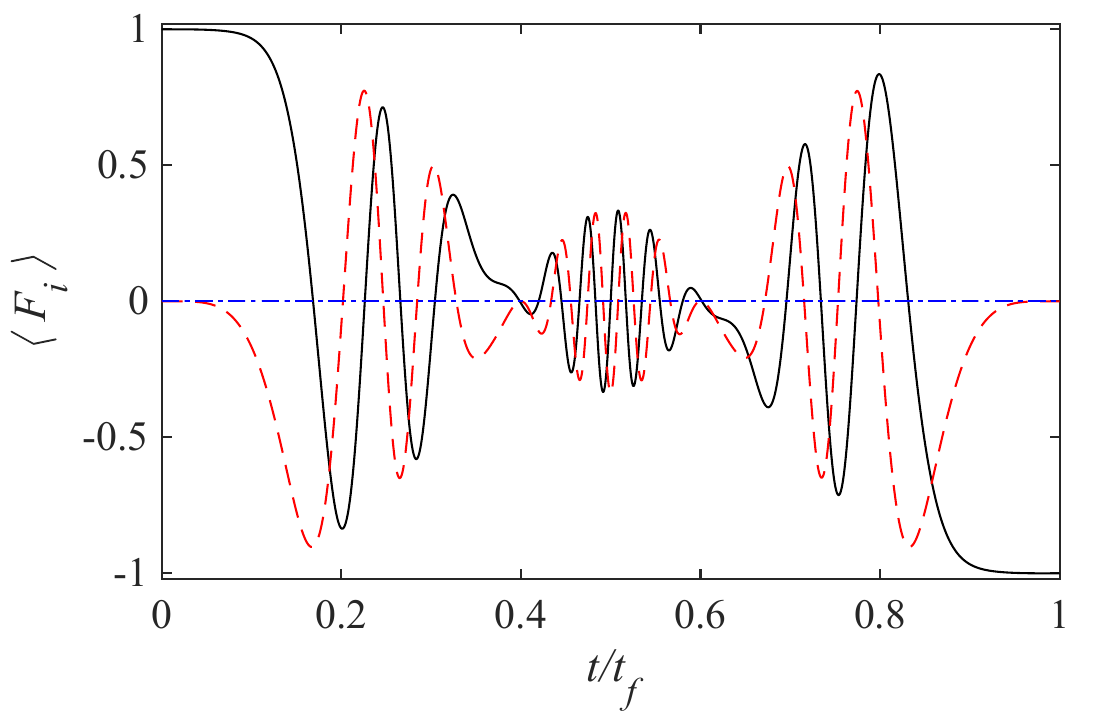}
\caption{Time evolution of spin components $\left\langle F_i\right\rangle$ during the fast transport, representing $\left\langle F_x\right\rangle$ (solid black line), $\left\langle F_y\right\rangle$ (dashed red line), 
and $\left\langle F_z\right\rangle$ (dotted-dash blue line). Parameters are the same as those in Fig. \ref{wavepacket}.}
\label{spindensity}
\end{figure}

Furthermore, we analyze the spin evolution using the reduced density matrix. Following Ref. \cite{mardonov2015dynamics}, the density matrix is defined as: 
$ 
\rho(t)=|\Psi(x, t)\rangle\langle\Psi(x, t)|$, 
where $\rho_{i j}(t)=\int \Psi_i(x, t) \Psi_j^*(x, t) d x \quad(i, j=1,0,-1)$. The reduced density matrix adheres to the wave function's normalization, $\operatorname{tr}(\rho)=\rho_{11}(t)+\rho_{22}(t)+\rho_{33}(t)=1$, and defines the spin component expectations as $\left\langle F_i\right\rangle=\operatorname{tr}\left(F_i \rho\right)(i=x, y, z)$. Fig. \ref{spindensity} illustrates the spin evolution over time. At the outset, with the spin parallel to the $x$-axis, the spin polarization's expectation value is $\left\langle F_x\right\rangle(0)=1$, transferring to $\left\langle F_x\right\rangle(t_f)=-1$ at the end time. Consequently, atoms are transported while the spin rotates around the $z$-axis and flips. Although the spin remains in a pure state initially and finally, non-adiabatic transport induces spin-dependent excitation, resulting in a mixed state in the spin subspace.

\subsection{Fast wave-packet splitting}

Now let us investigate the fast splitting of a wave packet with spin-orbit coupled spin-1 BECs via tunable SOC strength. For simplicity, we consider a scenario with a static harmonic trap, where $x_0(t)=0$, although including a moving trap is also possible.

The single particle Hamiltonian (\ref{h}) is reduced to $ H =  p_x^2/2m + m\omega^2 x^2/2 + \alpha(t) p_x F_z $. In this case,
the unitary transformation is nothing but $\mathcal{U}_s(t)$ with $\mathcal{U}_o(t)=1$. Utilizing Eq. (\ref{Psi}),  the wave function solution 
$
|\Psi (t) \rangle= \mathcal{U}^{\dag}_s(t) e^{- i H_0 t/\hbar}   \mathcal{U}_s(0) |\Psi(0) \rangle 
$,
can be expressed as: 
\begin{equation}
\label{coherent}
|\Psi (t) \rangle= e^{i \phi_{\alpha}} e^{\left[\beta(t) a^{\dagger}-\beta^*(t) a\right] F_z}  e^{- i H_0 t/\hbar}  |\Psi(0) \rangle,
\end{equation}
from which we introduce the displacement operator as
\begin{equation}
\hat{D}(\beta) = e^{\left[\beta(t) a^{\dagger}-\beta^*(t) a\right] F_z},
\end{equation}
with 
\begin{equation}
   \beta(t) = \sqrt{\frac{m\omega}{2 \hbar }} \left[\frac{\dot{\alpha}_c (t)}{\omega^2} - i\frac{\alpha_c (t)}{\omega} \right].
\end{equation}
Without loss of generality, we assume the initial wave function:
\begin{equation}
\label{splittingwavepacket}
|\Psi(x, 0)\rangle=\frac{ \sqrt{3}}{3}\left(\begin{array}{l}
1, 
1,
1
\end{array}\right)^{\mbox{T}} \otimes|\psi(x, 0)\rangle,
\end{equation}
where the wave function is the ground state of the stationary harmonic oscillator
\begin{equation}
\label{initial-splitting}
|\psi(x, 0)\rangle=\left(\frac{1}{\pi a^2}\right)^{1 / 4} \exp \left(-\frac{x^2}{2 a^2}\right).
\end{equation}
Since the coherent state is specific to the harmonic oscillator, we can determine the center-of-mass position of the atomic packet for the spin-up and spin-down components as:
\begin{equation}
\label{x}
\langle x_{\pm} (t) \rangle = \pm
\frac{1}{\sqrt{2}} a_0
[\beta(t)+\beta^*(t)]= \pm \frac{\dot{\alpha}_c(t)}{\omega^2},
\end{equation}
respectively. Additionally, the corresponding expectation value of momentum is given by:
\begin{equation}
\label{p}
\langle p_{\pm} (t) \rangle = \pm \frac{1}{i \sqrt{2}} \frac{\hbar}{a_0}[\beta(t)-\beta^*(t)] = \mp m \alpha_c(t),
\end{equation}
from which the velocity (\ref{velocity}) for each wave packet corresponding to spin-up and spin-down components can be obtained from Eq. (\ref{alpha}) as 
\begin{equation}
\label{velocityspin}
\langle v_{\pm} (t) \rangle = \mp [\alpha_c(t) - \alpha(t)] = \pm  \frac{\ddot{\alpha}_c(t)}{\omega^2}.
\end{equation}
This is consistent with the derivation of Eq. (\ref{x}) with respect to time.
With the expressions provided in Eq. (\ref{coherent}) and considering the initial state assumed in Eq. (\ref{splittingwavepacket}), we can achieve the final state with equal spin components at $t=t_f$, as depicted below:
\begin{align}
\label{final-splitting}
|\Psi_{i}\left(x, t_f\right)\rangle &= \frac{\sqrt{3}}{3}\left(\frac{1}{\pi a^2}\right)^{1 / 4} \exp \left[-\frac{(x^2-x^2_i)}{2 a^2}\right],
\end{align}
where $i=\{1,0,-1\}$, $x_{1}= d$ ($x_{-1}= -d$) represents the final positions of the two splitted wave packets with spin-up (spin-down) components, while $x_{0}= 0$ for the zero spin component.

\begin{figure}
\centering
\includegraphics[width=0.45\textwidth]{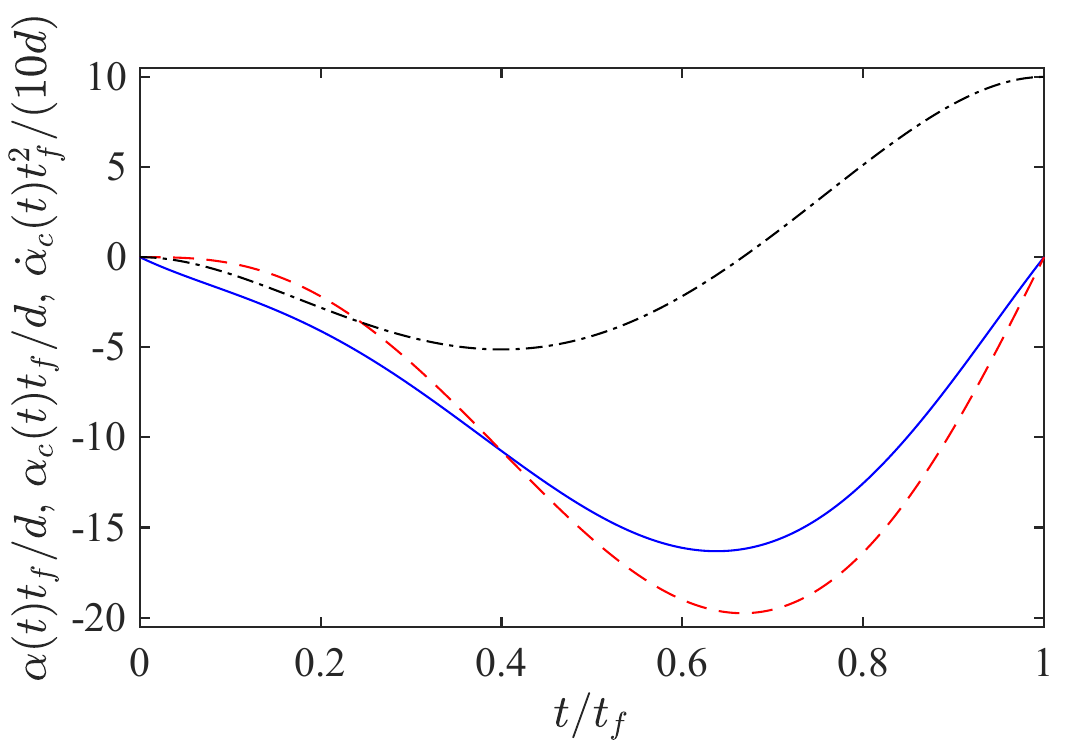}
\caption{Dependence of the SOC strength $\alpha(t) t_f/d$ (solid blue line) designed from $\alpha_c(t) t_f/d$ (dashed red line) and trajectory $\dot{\alpha}_c (t) t_f^2/(10 d)$ (dotted-dash black line) on time $t/t_f$. All quantities are dimensionless. Parameters: $t_f=10/\omega$ and d=$10$.}
\label{SOCsplitting}
\end{figure}

With these background and understanding, we can proceed to engineer inversely the time-dependent SOC strength to realize the desired fast splitting within the prescribed time $t_f$ and distance $d$. To do this, we set the following boundary conditions, 
\begin{equation}
\label{BC-3}
\begin{gathered}
\alpha_c(0)=0, \quad 
\dot{\alpha}_c(0)=0, \quad  \ddot{\alpha}_c(0)=0,   \\
\alpha_c(t_f)=0, 
\quad  \dot{\alpha}_c(t_f)= \omega^2 d, \quad  \ddot{\alpha}_c(t_f)=0.
\end{gathered}
\end{equation}
From Eqs. (\ref{velocity}), (\ref{x}) and (\ref{p}), the boundary conditions mentioned above suggest the final velocity for spin-up and spin-down components is zero, as the initial one. Along this, we simply choose the polynomial ansatz in the form $\alpha_{c}(t)=\sum_{i=0}^5 a_i t^i$ to satisfy the above boundary conditions, such that the function of $\alpha_c(t)$ is obtained as
\begin{equation}
\label{alphac}
  \alpha_c (t) = -d \omega^2 t_f(3s^5-7s^4+ 4s^3),
\end{equation}
which determines the trajectories of the splitted wave packets with spin-up and spin-down components through Eq. (\ref{x}). Then, the combination of Eqs. (\ref{alpha})
and (\ref{alphac}) finally gives
\begin{equation}
\alpha (t) = -\frac{d s}{t_f} \left[\omega^2 t^2_f(4s^2-7s^3+3s^4) + 12 (5s^2-7s+2) \right].
\end{equation}
As depicted in Fig. \ref{SOCsplitting}, the time-dependent SOC strength $\alpha(t)$ designed here, is compared with the trajectory of the splitted wave packets, which is determined by $\dot{\alpha}_{c}(t) /\omega^2$.

\begin{figure}
\centering
\includegraphics[width=0.45\textwidth]{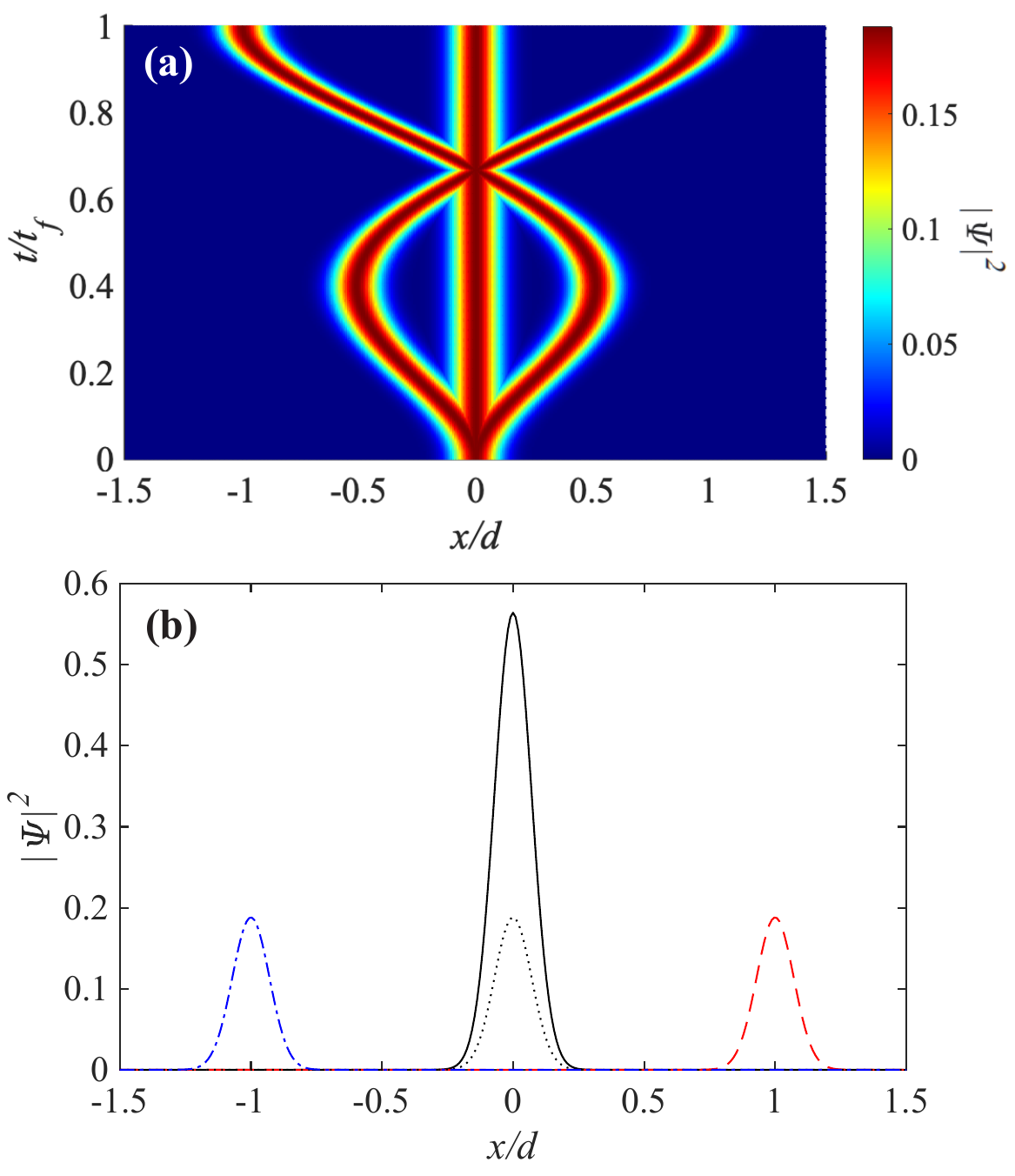}
\caption{(a) The contour map illustrates the propagation of three spinor wave packets during the fast splitting process designed via the inverse engineering method. In this process, the wave packets are finally separated by the distance $\pm d$ for the spin-up and spin-down components, while the zero spin component remains static. (b) The density distribution displays the initial total wave function $|\Psi(x,0)|^2$ (black solid line) and the density distribution of the three spin components $|\Psi_{1,0,-1}(x,t_f)|^2$ (red dashed line, black dotted line, and blue dotted-dash line) at $t=t_f$. The parameters are consistent with those used in Fig. \ref{SOCsplitting}.}
\label{splittingevolution}
\end{figure}

Figure \ref{splittingevolution}(a) illustrates the fast splitting process of spin-up and spin-down polarized wave packets, which are transformed to positions $\langle x_{\pm} (t_f) \rangle = \pm d$, fulfilling the boundary conditions (\ref{BC-3}). Meanwhile, the zero spin component remains static without any displacement. 
It is notable that the trajectories of the spin-up and spin-down components are symmetrically separated, a consequence of the opposing velocity of the two components, as denoted by Eq. (\ref{velocityspin}). To clearly depict the results of the wave packet splitting, the probability distributions of wave pacekts at the initial and final times are shown in Fig. \ref{splittingevolution}(b).

As a matter of fact, the utilization of time-dependent SOC serves a similar function to spin-dependent forces in the controlled splitting of atomic wave packets \cite{Muga2020}. Our approach offers alternative promising avenues for enhancing atomic interferometry, providing precise manipulation and control over atomic states to enable high-resolution measurements. Through the inverse engineering of SOC, tailored atomic wave packets with specific spatial and spin characteristics can be generated, potentially enhancing the sensitivity and accuracy of interferometric analyses. Further investigation into this approach, even with moving trap, is worthwhile to explore its nuances and potential applications.

\section{COMPARISON WITH A SIMPLE CASE OF CONSTANT SOC AND VELOCITY}

\label{constant}

While the time-dependent SOC and trap trajectory provide intricate control over the dynamics, the simpler case of constant SOC and velocity of trap provides a baseline for comparison. Constant SOC, albeit less flexible in its manipulation, still influences the motion and spin dynamics of atoms.  

Firstly, to illustrate the advantage of the inverse engineering method, we consider a scenario with constant SOC strength, denoted as $\alpha$, and constant velocity, represented by $d/t_f$. When $t_f$ is sufficiently large, the adiabatic condition is fulfilled, ensuring that $\dot{x}_c$ and $\ddot{x}_c$ are negligible, leading to $x_c(t) = x_0(t)$. By substituting the constant SOC strength $\alpha$ into Eq. \eqref{alpha}, we have
\begin{equation}
\label{costantalpha}
\alpha_c(t)=\alpha[1-\cos(\omega t)],
\end{equation}
with the initial boundary conditions $\alpha_c(0) = \dot\alpha_c(0) = 0$. 
When $\omega t_f = 2k\pi$, $(k = 1,2,3,...)$, the boundary conditions $ \alpha_c(t_f) = \dot\alpha_c(t_f) = 0$  are fulfilled. Moreover, the phase factor \eqref{phi} is obtained as
\begin{equation}
\label{phase}
\phi (t_f)=-\frac{m d \alpha}{\hbar\omega t_f}[\sin(\omega t_f)-\omega t_f \cos(\omega t_f)],
\end{equation}
by using $\dot\alpha_c(t)=\alpha \omega \sin(\omega t)$ and $x_0(t)=dt/t_f$. 
When $\omega t_f = 2k\pi$, the phase factor  becomes $\phi (t_f)=d/\lambda_{so}$, where $\lambda_{so}=\hbar/(m\alpha)$, representing the SOC length. By imposing $\phi (t_f)=\pi$, we get the characteristic length for spin flip $d_{sp}=\pi \lambda_{so}$, resulting in the spin-flip time, $t_{sp}=d_{sp}t_f/d$.

\begin{figure}
\centering
\includegraphics[width=0.45\textwidth]{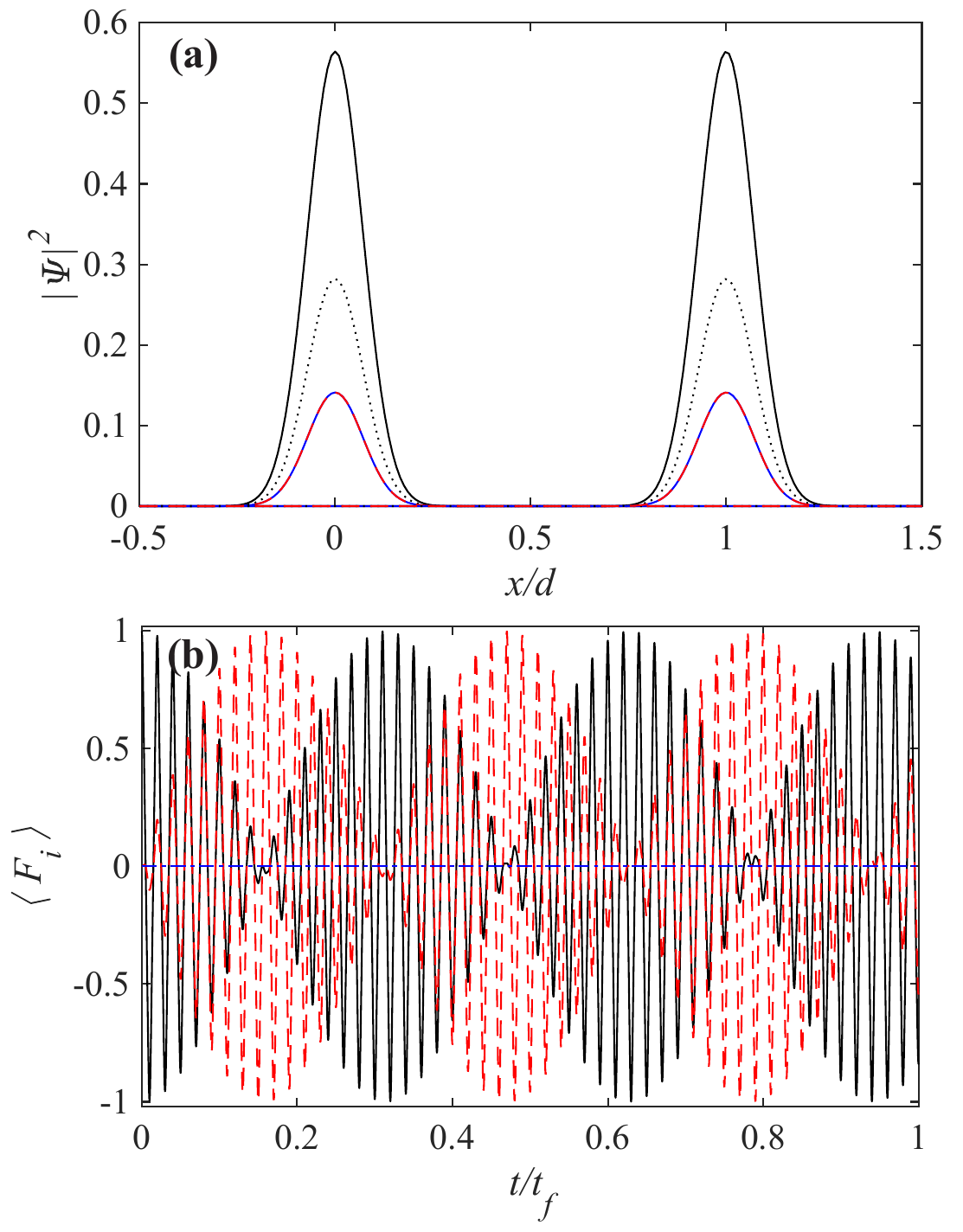}
\caption{(a) The density distribution of the total wave function $|\Psi(x,t)|^2$ (black solid line) at $t=0$ and $t=t_f$, along with the density distribution of the three spin components, denoted by $|\Psi_{1,0,-1}(x,t)|^2$ (denoted by blue solid line, black dotted line and red dashed line). (b) Time evolution of spin components $\left\langle F_i\right\rangle$ in the presence of constant SOC strength during the adiabatic transport, 
where $\left\langle F_x\right\rangle$ (solid black line), $\left\langle F_y\right\rangle$ (red dashed line) , and $\left\langle F_z\right\rangle$ (blue dotted-dash line) are  depicted respectively. Parameters: $\alpha=1$, $t_f=200\pi/\omega$, and $d=10$.}
\label{costant}
\end{figure}

Figure \ref{costant}(a) illustrates that the spin-orbit coupled spin-1 wave packet of atoms can be transported from $x_0 = 0$ to $x_0 = d$ when $t_f= 200\pi/\omega$, a duration long enough to satisfy the adiabatic criterion. In this scenario, $t_f$ is an integer multiple of $2\pi/\omega$, resulting in $\alpha_c(t_f) = \dot\alpha_c(t_f) = 0$, ensuring no final excitation of the orbital motion and an exact displacement of $d$ for the wave function. Concurrently, the spin dynamics is governed by the characteristic length $d_{sp}$, indicating the spin flip can be achieved if the transport distance is $d_{sp}$ or an odd multiple of $d_{sp}$. As depicted in Fig. \ref{costant}(b), when $d = 10 $ and $\alpha = 1 $, the spin flip time is $t_{sp} = 0.314t_f$. In this case, $d\neq(2k-1)d_{sp}$, hence the final spin state is not the eigenstate of $F_x$, preventing complete spin flipping. However, at the final moment, the wave function of the three spin components coincides with the same displacement $d$. Consequently, the atomic wave pacekt can be adiabatically transported from $x_0=0$ to $x_0=(2k-1)d_{sp}$, while achieving spin flip simultaneously. It's worth noting that, in the adiabatic approximation, the characteristic length for spin flip depends solely on the SOC strength, not the transport velocity. Adiabatic transport with constant SOC strength and velocity has a comparable effect on orbital and spin dynamics as the inverse engineering method, albeit requiring a specific longer time and limited to relatively small final positions.

Secondly, we examine the splitting process with constant SOC strength 
$\alpha$ for completeness. Again, we have $\alpha_c(t)=\alpha[1-\cos(\omega t)]$ and $x_0(t)= x_c(t)=0$. Based on Eq. \eqref{x}, we derive the center-of-mass position of the atomic wave packets for the spin-up and spin-down components as:
\begin{equation}
\label{conx}
\langle x_{\pm} (t_f) \rangle = \pm \frac{\dot{\alpha}_c(t_f)}{\omega^2}=\pm\frac{\alpha \sin(\omega t_f)}{\omega}.
\end{equation}
Also, the velocity for each wave packet can be obtained from \eqref{velocityspin} as:
\begin{equation}\label{velocitycon}
\langle v_{\pm} (t_f) \rangle  = \pm  \frac{\ddot{\alpha}_c(t_f)}{\omega^2}=\pm \alpha \cos(\omega t_f).
\end{equation}
Thus, Eq. \eqref{conx} illustrates that the center-of-mass position of the wave packets with the spin components varies with time $t_f$. Notably, when $\omega t_f = 2k\pi$ $ (k = 1,2,3,...)$, $\langle x_{\pm} (t_f) \rangle=0 $, and the atomic packet can not split into three components. Additionally, to achieve stable wave packets with three components, the final velocity for spin-up and spin-down components must be zero. According to Eq. \eqref{velocitycon}, this condition requires
$\omega t_f = \pi/2+k\pi$ $ (k = 1,2,3,...)$ to be satisfied. Based on this analysis, although constant SOC can achieve stable wave packet splitting, it provides only specific values of $t_f$. In this sense, the inverse engineering method offers more flexibility and convenience.

\section{Numerical simulation}

\label{nonlinear}

In this section, we shall explore the interatomic collisional interactions on the transport and splitting processes, designed by our inverse engineering method. When nonlinear interactions exist between atoms, the  mean-field Hamiltonian in Eq. (\ref{GP}) can be expressed as \cite{SuotangPRA2020}:
\begin{equation}
H_{int} = 
\begin{pmatrix}
\Gamma_{1} & 0 & 0 \\
0& \Gamma_{0} & 0\\
0&0 & \Gamma_{-1} \\
\end{pmatrix},
\end{equation}
where $\Gamma_{\pm} = (c_0+c_2)(|\psi_{1}|^2 + |\psi_{0}|^2 + |\psi_{-1}|^2)- 2 c_2 |\psi_{\mp}|^2$ and
$\Gamma_{0}  =(c_0+c_2)(|\psi_{1}|^2 + |\psi_{0}|^2 + |\psi_{-1}|^2)-  c_2 |\psi_{0}|^2$.
The nonlinearities are determined by density-density interactions with the coefficient $c_0$ and spin-spin interactions with the coefficient $c_2$. Typically, for $^{87}$Rb atoms, one can select parameters such as $c_0 = 0.05$ (in the unit of $\sqrt{m/\hbar^2 \omega}$), along with a ratio of $c_2/c_0 = -0.005$ in the literature \cite{SuotangPRA2020}. Moreover, in another study \cite{PhysRevA.108.033317}, different values of $c_0= 0.25$ and $c_2=-0.001$ are used.
As a matter of fact, the strength of these interactions can be adjusted using Feshbach resonance techniques. Hence, we intend to  investigate the stability of our STA protocols for transport and splitting within the region of nonlinear interaction.

\begin{figure}
\centering
\includegraphics[width=0.45\textwidth]{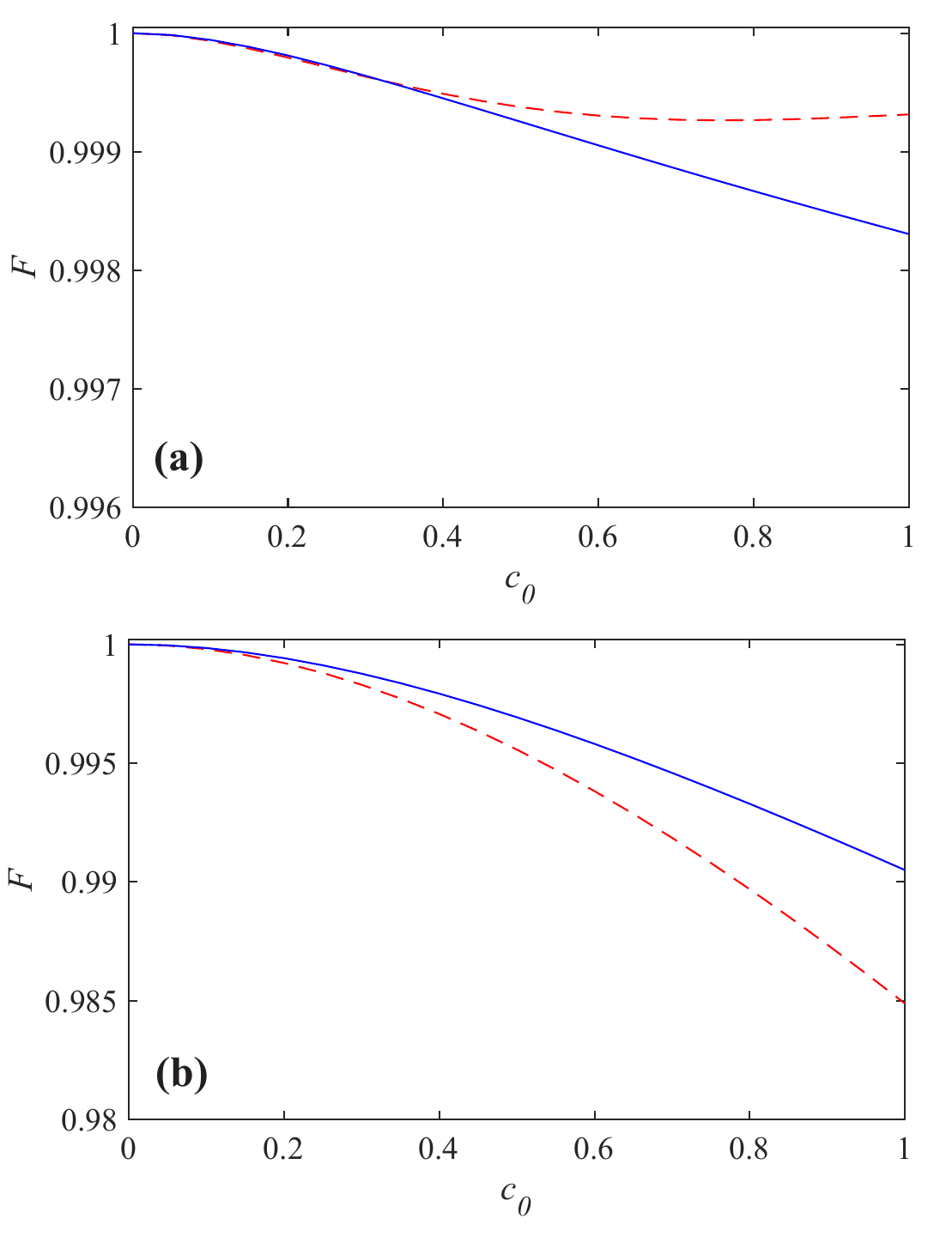}
\caption{The fidelity of (a) fast transport with spin flip and (b) splitting is compared against the nonlinearity parameter $c_0$, where the Gaussian initial states are used. In (a) the time-dependent SOC and moving trap trajectories are designed according to Fig. \ref{x&alpha}, and in (b) the time-dependent SOC is designed according to Fig. \ref{SOCsplitting}. Parameters: $c_2/c_0 = -0.005$ (red dashed line), $c_2/c_0= -1$ (blue solid line) and other parameters consistent with those in Figs. \ref{x&alpha} and \ref{SOCsplitting}, respectively.}
\label{fidelity}
\end{figure}

To quantify the stability against the nonlinear interaction, we define the fidelity as:
\begin{equation}
\label{42}
F=\left|\langle\Psi (x,t_f) \mid \tilde{\Psi}\left(x, t_f\right)\rangle\right|^2,
\end{equation}
where $|\Psi (x,t_f) \rangle$ represents the target state at $t=t_f$ obtained through inverse engineering, and $| \tilde{\Psi} (x,t_f) \rangle$ 
is the exact evolution of the wave packet obtained using the split-operator method, particularly for spinor BECs  \cite{Chaves2015,gawryluk2018unified}.

\begin{figure}
\centering
\includegraphics[width=0.45\textwidth]{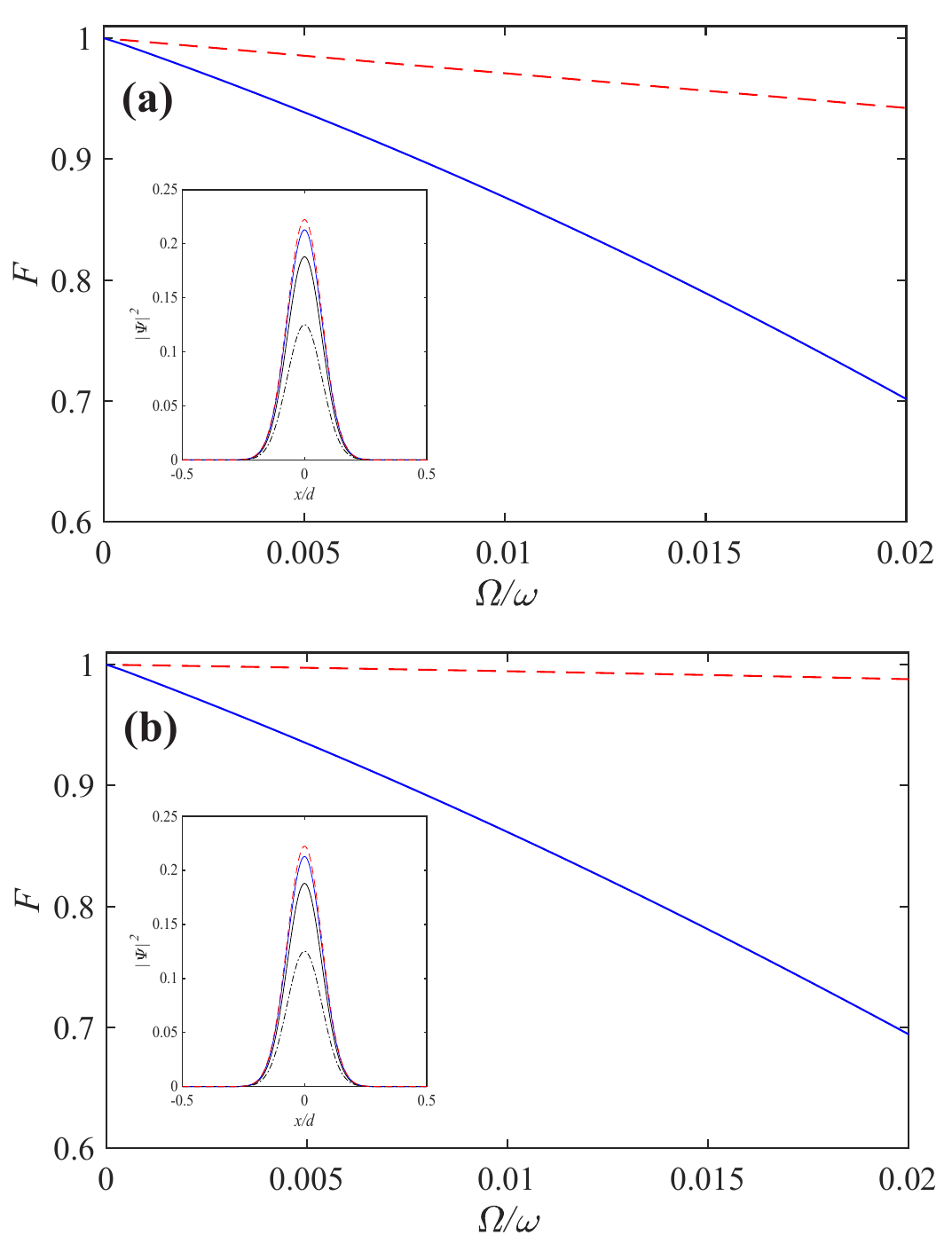}
\caption{ The fidelity of (a) fast transport with spin flip and (b) splitting is compared against the transverse potential $\Omega$, where the Gaussian initial state (red dashed line) and the stationary state obtained from the imaginary time evolution (blue solid line) are used.  In (a) the time-dependent SOC and moving trap trajectories are designed according to Fig. \ref{x&alpha}, and in (b) the time-dependent SOC is designed according to Fig. \ref{SOCsplitting}. The inset illustrates the difference between the Gaussian ground state of the harmonic trap (black solid line) and the stationary state $|\Psi_{1,0,-1}(x,0)|^2$ obtained from the imaginary time evolution (blue solid line, black dotted-dash line, and red dashed line), for example, at $\Omega/\omega=0.02$. Parameters: $c_0=0.05$, $c_2/c_0 = -0.005$, and other parameters consistent with those in Figs. \ref{x&alpha} and \ref{SOCsplitting}, respectively.}
\label{transversepotential}
\end{figure}

In Fig. \ref{fidelity}(a), the fidelity of the rapid transport with spin flip is illustrated using the designed time-dependent SOC strength $\alpha(t)$ and trap trajectory $x_0(t)$ as shown in Fig. \ref{x&alpha}. Following the concept of STA, the initial and final states are assumed by Eqs. (\ref{initial}) and (\ref{final}). However, the exact final state, calculated from the split-operator method, slightly deviates from the target state due to interatomic interactions. With the influence of interatomic interaction, the final state obtained from numerical simulation is different from the target one. Consequently, the fidelity in terms of the nonlinearity parameter $c_0$ deteriorates, especially as we increase the ratio $c_2/c_0$ from $-0.005$ to $-1$.
In Fig. \ref{fidelity} (b), we further elucidate the fidelity of fast splitting using the STA protocol with the time-dependent SOC strength $\alpha(t)$, depicted in Fig. \ref{SOCsplitting}. Once again, the initial and final states are determined by Eqs. (\ref{initial-splitting}) and (\ref{final-splitting}). Again, we observe the fidelity decreases with increasing the nonlinear parameter $c_0$.
 However, when altering the ratio $c_2/c_0$ from $-0.005$ to $-1$, this fidelity behavior differs from that observed in the transport case, owing to differences in the probability of spinor wave functions. Nonetheless, Fig. \ref{fidelity} highlights that the fidelity remains reasonably high when $c_0=1$ and $c_2/c_0=-1$. This underscores the importance of mitigating the effects of atomic interactions in the context of SOC BECs for precise control and manipulation of atomic states, utilizing our analytical STA presented before. In addition, for even larger nonlinear interaction $c_0 \simeq 5$ and $c_2/c_0=-1$ (not depicted in Fig. \ref{fidelity}), the fidelity is far from perfect. In such cases, we have to emphasize that alternative methods for designing STA, including hydrodynamic formalism \cite{StringarihydroPRL} and variational approximation \cite{Tangyou}, become necessary.

 Finally, we  attempt to  access the stability concerning the transverse  potential $\Omega$ described in Eq. (\ref{h}). We note that when $\Omega=0$, the conservation of the spin component  $F_z$ is observed. Consequently, the orbit motion corresponding to the center of mass remains unaffacted by the detuning $\delta$, as indicated in Eq. (\ref{velocity}). Hence, we set $\delta$ to zero for this case.  However, the appearance of the transverse potential $\Omega$ breaks the conservation of $F_z$, altering the dynamics. Consequently, the fidelity decreases with the Zeeman potential in both transport and splitting processes, as illustrated in Fig. \ref{transversepotential}. Furthermore, to prepare the initial state with the spinor component along $F_x$ in a realistic experiment, we need to switch on the Zeeman potential since the SOC strength is initially zero. By gradually increasing the Zeeman potential, the energy spectrum corresponding to BECs changes from parabolic to non-parabolic energy dispersion, leading to changes in the eigenstates accordingly.  In insets of Fig. \ref{transversepotential}, we showcase the stationary state through imaginary time evolution, which differs from the Gaussian ground state of the harmonic trap. Naturally, the fidelity worsens as a consequence in this scenario. In the further work, the SOC strength and trap trajectory can be optimized with respect to the transverse  potential $\Omega$, or variational dynamics \cite{MotionPRAVA} combined with inverse engineering may be employed to address this issue.

\section{CONCLUSION}
\label{conclusion}

In summary, we have delved into the dynamics of spin-orbit-coupled spin-1 BECs, specifically focusing on their fast transport with spin flip and efficient splitting using the inverse engineering method rooted in STA.
Initially, we have achieved fast transport of spin-orbit-coupled spin-1 BECs with spin flip within short time. This involved designing the potential position and time-dependent SOC strength using the inverse engineering method, with careful consideration of boundary conditions to ensure simultaneous rapid transport and spin flip. Subsequently, we have extended this method to achieve rapid splitting, generating spin-dependent coherent states within a short time scale in a static harmonic potential. Through rigorous numerical calculations, we have confirmed the coherence of the obtained states, validating the effectiveness of our proposed method. Moreover, we have conducted a comparative analysis between our proposed method and simple (adiabatic) case of constant SOC strength and/or trap velocity. Despite the challenges in controlling time-dependent SOC and moving potential, the inverse engineering method have demonstrated greater flexibility and efficiency, enabling faster and more precise control over atomic states. Additionally, we have examined the influence of nonlinear interactions described by the GPE on the transport and splitting processes. Despite the presence of atomic interactions, our proposed method remained effective, even with slightly different initial states in presence of transverse potential. 

However, several challenges and open questions remain. Experimental realization of the designed time-dependent potentials and SOC strengths somehow poses technical challenges \cite{SpielmanPRLtunableSOC,luo2016tunable}. In addition, understanding the effects of environmental noise and systematic errors on the fidelity, scalability to larger systems with multiple interacting particles, and extension to other quantum systems beyond spin-orbit coupled spin-1 BECs are areas that require further investigation. For instance, the SOC strength and trap trajectory can be optimized with respect to systematic errors or noise by using optimal control theory \cite{Chenpra11,Lupra} and machine learning as well \cite{Dingpra}.  In a word, we have proposed to manipulate cold atoms in the ground state effectively within this framework, facilitating rapid transport \cite{Erikpra83,Qipra} and the creation of non-classical states \cite{Noripra81,Clerkpra96}. We hope the insights gained from our study have potential applications in quantum information processing and quantum metrology.

\begin{acknowledgements}
This work is supported by NSFC (Grant No. 12075145 and 12211540002), STCSM (Grant No. 2019SHZDZX01-ZX04), and the Innovation Program for Quantum Science and Technology (Grant No. 2021ZD0302302), the Basque Government through Grant No. IT1470-22, the project grant PID2021-126273NB-I00 funded by MCIN/AEI/10.13039/501100011033, by ``ERDFA way of making Europe",  ``ERDF Invest in your Future",  and Nanoscale NMR and complex systems (PID2021-126694NB-C21).  X.C. acknowledges ``Ayudas para contratos Ram\'on y Cajal'' 2015-2020 (RYC-2017-22482).
\end{acknowledgements}


\end{document}